# Statistical Analysis of Behavioral Intention Towards Private Umbilical Cord Blood Banking

Sonia Ghayem[1] and Reza Foudazi[2]

[1] Islamic Azad University- Center Tehran Branch, Tehran, Iran
`soniaghayem@gmail.com`
[2] Islamic Azad University- South Tehran Branch, Tehran, Iran
`rezaf1989@gmail.com`

**Abstract.** This paper proposes a conceptual framework to identify the key dimensions affecting behavioral intention to bank umbilical cord blood in Iran. We examine the impact of awareness, reference group, usability, disease history, and price on perceived risk and behavioral intention to use umbilical cord blood banking service. To evaluate the proposed model of umbilical cord blood banking behavioral intention and test our hypotheses, we apply exploratory field research. We use a five-point Likert scale to form a questionnaire to collect the data. The model is estimated with a sample of 242 Royan cord blood bank customers in Tehran. We use Pearson correlation and structural equation modeling to analyze the structural relationships between research variables, perceived risk, and behavioral intention. This research gives novelty on behavioral intention determinants in private umbilical cord blood banking, which adds value to literature and future managerial practices. Results show that usability is the primary determinant in cord blood banking.

## 1 Introduction

Riding on a trend that the numbers of cord blood banks are growing in the last decade (Lim, I. J., and Phan, T. T., 2014), it is envisaged that Iran is looking for a new era of promoting this new technology. In view of the fact that Iran is in the top 10 of countries in the world that produce, culture, and freeze human embryonic stem cells (Shaiegan, 2010), it is necessary to undergo research that may assist in explaining the factors influencing behavioral intention in cord blood banking and subsequently hasten the process of commercialization.

The cloning of Dolly the sheep and the multilineage differentiation of human embryonic stem cells have led to improvements in regenerative medicine (Wilmut *et al.*, 2007). Umbilical cord blood (UCB) is one of the most absorbing bio-objects in medical applications (Brand *et al.*, 2008). Privately banked UCB might offer treatment for regenerative medicine, Parkinson's, or diabetes (Goodarzi *et al*., 2015). The use of UCB stem cells has no concern with an embryo's destruction. Instead, it is effective in research and therapeutic procedures, which do not involve any moral and ethical concerns (Yang *et al.*, 2003). Umbilical cord blood banks store the newborn's umbilical cord blood stem cells for the family's private use. These stem cells will be used in regenerative medicine. Regenerative medicine suggests the process of restoring or reengineering the structure and function of damaged human cells, tissues, and organs to establish normal function (Armson *et al.,* 2015). With the exception of countries such as Italy and Spain, the privately run cord blood banks are now operating in almost all parts of the world (Sullivan, 2008). One reason for worldwide private cord blood banking is the interest of parents in giving their children biological insurance in circumstances when they develop different health problems in the future (Martin *et al.*, 2008). In countries such as Iran, the broad perceptions and ethical concepts of stem cells being unproblematic can open the doors of growth and develop a sense of belief in the donors as well as the common individuals so that they can trust cord blood banks in a better way.

Although several studies revolve around UCB banking points of argument (Gluckman, 2009; Fasouliotis and Schenker, 2000; Kline and Bertolone, 1998) remarkably, no research has examined the key contributing factors in private cord blood banking in the marketing-oriented literature review. Seeing this gap in previous studies, we study the factors motivating people to stock UCB.

Optimization methods use sophisticated analyzes to predict the outcome (see, for example, Mohamadi and Bahrini (2020); Ghassemi (2019); Peykani et al. (2019); Ghassemi et al. (2017); Esmaeili et al. (2015)) and may fail to solve the issue, as simulating a real problem such as in the UCB banking is not easy. Instead, in this research, we used a conceptual statistical model of UCB banking behavioral intention and tested hypotheses that are created from the model and applied field exploratory research.

The research draws on the concept of perceived risk to propose a conceptual model to address the hypothesized relationships. We propose a conceptual framework to point out whether UCB awareness, reference group, usability, disease history, and price may affect the perceived risk and accordingly behavioral intention to bank UCB stem cells for personal use. The variables were selected based on an extensive review of the literature (Yang *et al*., 2003). The contribution of this study would be a valuable one because the sample population of the study is already composed of private cord blood bank customers, whereas most of the past research studies that have offered insights into this issue revolve around pregnant women at hospitals (Fox *et al*., 2007; Yang *et al*., 2003; Katz *et al.*, 2011).

## 2     Theoretical Background and Hypotheses

In the profit-oriented industry of private UCB banking, patients are considered as consumers, and consumer behavior entails considerable risks and requires prudent decision making because of the uncertainty of the outcome (Fanoodi, 2019). The theory of perceived risk was first introduced by Bauer (1960) in the field of consumer behavior. Since 1960 (Cox, 1967), this theory was considerably applied in many stud-ies, and the impact of perceived risk in consumer behavior has been affirmed globally. In the case of UCB banking offering state-of-the-art technology, perceived risk was anticipated to play a role. The higher the risk perceived, the fewer consumers are inclined to pay for UCB banking.

## 2.1   UCB awareness

Awareness indicates the level of consumers' understanding, recognition, and recall (Aaker, 1991). UCB awareness is the level of experience and knowledge of UCB banking and its therapeutic applications (Yang et al., 2003). It is essential to remark that several studies have highlighted the effect of awareness on behavioral intention to use a service (Peykani et al., 2020; Yang et al., 2003; Homburg et al., 2010; Dowling and Stealin, 1994). According to Katz et al. (2011), about 92% of all pregnant women had a lack of knowledge toward UCB banking and showed their eagerness to be informed on UCB applications (80% in the UK to 98% in Italy). The findings of Fox et al. (2007), Yang et al. (2003), and Bhundari et al. (2017) also have indicated the noticeably poor understanding of pregnant women about the applications and the use of UCB stem cells in regenerative medicine. Besides, several studies also reported a lack of knowledge among healthcare professionals, which hinders UCB banking (Moustafa and Younes 2015, Hatzistilli et al. 2014). Many studies have tested and obtained support for the knowledge and awareness impact on customers' perceived risk and, consequently, on behavioral intention (see Sharp (1995) and Keller (1998), for example).

Drawing from the literature, when the level of UCB banking awareness is upgraded, we can establish the following hypothesis because high awareness reduces consumer risks:

$H_1$: UCB banking awareness has a negative effect on perceived risk.

## 2.2   Reference group

Based on Park and Lessig (1977), a reference group is a group where the individuals hold their attitudes in conformity to reduce the uncertainty about the outcome. Fishbein and Ajzen (1975) believe that an individual's intention to purchase a product or use a service is affected by the community and its people. Numerous studies demonstrated that family members' and friends' reference groups could influence behavioral intentions (Turner, 1991; Whittler and Spira, 2002). As an example of brand selection, see the studies in Moutinho (1987); Bearden and Etzel (1982); Moschis (1976); Park and Lessig (1977). Barkhordari et al. (2019) and Venkatesh and Davis (2000) have proven the influence of social factors on willingness to use new technology.

A high-risk level imbues the UCB banking decision-making process because the benefit is not received at the time of investment. The results of Katz et al. (2011) reveal that the majority of pregnant women (91%) believe that fathers need to be involved in the decision related to bank UCB stem cells (87% in the United Kingdom to 93% in Spain). This shows that fathers remain the basis in the decision-making process, and they should be enlightened regarding the innovative service of UCB banking. Thus, we can establish the following hypothesis:

$H_2$: A reference group has a negative effect on UCB banking perceived risk.

## 2.3 Usability

Usability is the degree to which a person believes that using a particular system would enhance their performance. The UCB banking context is the effectiveness and satisfaction level of users who accept bank UCB stem cells for personal use and result in a favorable outcome. The technology acceptance model is widely used by scholars to explore technology adoption and usage (Koufaris, 2002; Moon and Kim, 2001). Fox et al. (2007) demonstrated that the women bank their child's UCB stem cells as biological insurance (83%). However, UCB stem cells' private storage remains controversial; the therapeutic use of UCB stem cells for transplantation is rising (Armson et al., 2015). According to Fox et al. (2007), private UCB banking's feasibility is reduced because of the high degree of uncertainty about the banked UCB stem cells' future benefits. The probability that a child will ever use their banked cord blood ranges from 1:1000 to 1:200000, so people should not take the risk (Fisk et al., 2005).

$H_3$: The usability of UCB has a negative effect on perceived risk.

## 2.4 Price

Cord blood banking is a procedure that requires fee payment in two phases; enrolment, collection, and storage are the initial charges that cover the first one-year, and the second is the annual storage fee. The initial amount varies from one bank to another, depending on the storage period, from $1000 to $2000. After the first year, storage fees range from $100 to $150 (Yang et al., 2003; Martin et al., 2008). The results of a study in France, Germany, Italy, Spain, and the UK regarding pregnant women and their understanding and feeling of cord blood stem cells and UCB banking showed that due to the costs, more than half of them would not store cord blood in private banks (Katz et al. (2011)). There was a relatively positive response to the view between the buying intentions and the cost of operation being low.

$H_4$: Price has a positive effect on UCB banking perceived risk.

## 2.5 Disease history

In the healthcare industry, a family disease history consists of information about disorders from which the patient's direct blood relatives have suffered. When a healthy child with an ill brother is born, and his umbilical cord is preserved, the stem cells are beneficial. It is an accepted fact that UCB storage in families with high risk makes UCB banking cost-effective. Fox et al. (2007) revealed that of his research population, only 5% had a history of a family disease, whereas only 2% had a child who had the disease. Fisk et al. (2005) also found out that a family with no family disease history does not need banked cord blood. Different researches have different views on UCB banking and the disease history influence. Generally speaking, if a family has a history of some diseases such as aplastic anemia, thalassemia, leukemia, and other similar disorders, the need for stem cell transplantation could be higher than the rest of the population. Therefore, there is an increased use of UCB banking among the people with a disease, which has reduced the UCB banking perceived risk. The perception of UCB banking not meeting the consumers' expectations among the high-risk families will be reduced (Fisk et al., 2005). Thus, we hypothesize the following:

$H_5$: Disease history has a negative effect on UCB banking perceived risk.

## 2.6 Perceived risk

The amount and nature of risk perceived by consumers in considering a purchase is the definition of perceived risk in business. Researchers such as Stone and Gronhaug (1993) define perceived risk in terms of the uncertainty and results, whereas others (Roselius, 1971; Bettman, 1973; Dowling and Staelin, 1994) consider it as the decision-making importance and any likelihood of negative results that come with the decision. The doubt in technology increases the risks associated with cord blood banking and its small chances of being put to use. According to Littler and Melanthiou (2006), there is a high risk among consumers during the early stages of introducing innovation. The consumers' high-risk perception should be reduced by taking practical actions to attract extensive expert attention and advance the UCB banking service development. This will, therefore, be of great practical importance. The consumers' adoption and willingness to UCB banking are highly dependent on their perceived risk, whereas its usability is the leading cause of the consumers' perceived risk (Fisk et al., 2005). Therefore, improving the UCB usability and its perceived risk will positively support the adoption of UCB banking.

## 2.7 Behavioral intention

A consumer's willingness to buy or use a service in the future is called behavioral intention (Fishbein and Ajzen, 1975). This is emphasized continuously in the literature that if the perceived risk is high, there will be a lower proneness to use a service (Lin, 2008; Mitchell, 1999; Dowling and Staelin, 1994). In the theory of consumers' perceived risk, it is shown that when the consumers were deciding on buying a product, their level of perception would eventually affect their decisions (Taylor, 1974). Also, different risk levels would create various changes with additional services and persons. As a result, perceived risk had considerable influence on the consumers' behavioral intention. Through the above discussion, we can establish the following hypothesis:

$H_6$: There is a negative relationship between perceived risk and intention to bank UCB.

Through the inference of the above literature and hypotheses, we can propose this research's conceptual model as Fig. 1.

We map marketing concerns about private cord blood banking from those already banking UCB stem cells of their children in Royan cord blood bank. What is motivat-ing people to bank UCB stem cells for personal use, and what is their effect on perceived risk? By arraying these research questions, we identify the plausible conceptual model. In some respects, this big picture is simplistic, but it is usefully showing the different factors affecting perceived risk in UCB banking.

The usefulness of the proposed model lies in its ability to provide a broad view of perceived risk in UCB banking for personal use and highlight the factors affecting perceived risk to take into account in developing private cord blood banks.

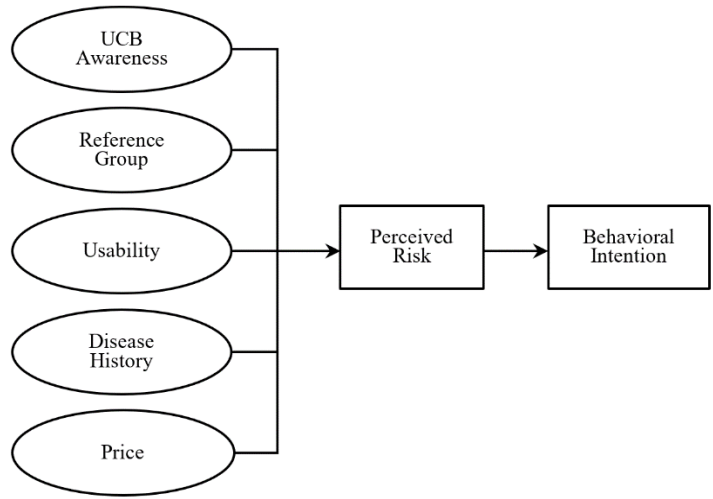

**Fig. 1.** A conceptual model.

## 3  Research Methodology

The study aims to determine causal relationships between variables. Based on the purpose of the study and the data collection method, the research is descriptive and correlational and specifically used SEM to analyze the structural relationship between research variables. UCB awareness, reference group, usability, disease history, and price are exogenous latent variables in the proposed conceptual model, where perceived risk and behavioral intention are endogenous latent variables. Despite this, UCB awareness, reference group, usability, disease history, and price can be described as independent variables where the perceived risk is an intervening variable, and behavioral intention is the dependent variable.

The survey population consisted of people who already decided to store UCB in Tehran's Royan cord blood bank. The Royan Institute for Reproductive Biomedicine has been chosen as one of the well-versed and leading stem cell institutions in Asia. It has so far been able to store more than 35000 stem cell samples in its UCB bank. Random sampling is employed. The random selection of the sample units is based on the different segments of a population, which has high information, and research characteristics of interest (e.g., Bosch and Wildner, 2003; Guarte and Barrios, 2006).

Of 450 questionnaires administered, 316 returned in Royan. We discarded 74 incomplete questionnaires, and in the end, we produced 242 valid questionnaires. The valid return rate was around 54%. Each variable is evaluated and measured with 5-point Likert-type scales with anchors of 1 to indicate "strongly disagree" and 5 to indicate "strongly agree." To check how the drivers of UCB banking affect perceived risk and behavioral intention as well as attitudes toward private UCB banking, the questionnaires consisted of two parts: the first part included demographic questions (sex, age, education, and annual income level); the second part focused on knowledge about UCB and examined the preference of participants regarding UCB private banking and the reasons for their choice. We adopted all of Tanaka (2004) measures and research conducted by Yang et al. (2003). We made the necessary modifications to address the particular context of cord blood banking in Iran.

## 4 Analysis and Results

### 4.1 Confirmatory Factor Analysis (CFA)

In order to assess reliability, we pretested 40 questionnaires. The Cronbach α reliability coefficient values are shown in Table 1. Only questions 4 and 7 were filtered out.

**Table 1.** Reliability of each measurement variable.

| Variables | Number of measurement variables | Cronbach α | Excluded questions | Cronbach α after excluding |
|---|---|---|---|---|
| Awareness | 3 | 0.77 | Quest. 4 | 0.78 |
| Price | 2 | 0.63 | Quest. 7 | 0.72 |
| Reference group | 3 | 0.72 | | |
| Usability | 5 | 0.73 | | |
| Disease history | 3 | 0.81 | | |
| Perceived risk | 3 | 0.76 | | |
| Behavioral intention | 3 | 0.75 | | |

Since all Cronbach α values were greater than 0.7, all 7 constructs were deemed acceptable (Nunnally, 1978). To test the validity, we used factor analysis, which is a statistical technique frequently used in social studies. Exploratory factor analysis of the exogenous and endogenous variables revealed the values of Kaiser-Meyer-Olkin (KMO) 0.84 and 0.82, respectively, which shows the adequacy of sample size for research variables. Furthermore, the significant coefficient of the Bartlett test was zero (less than 0.50). To check the validity of the measurement model, we used the confirmatory factor analysis. The results revealed that all factor loadings were greater than 0.3, which indicates a significant level of convergent validity.

### 4.2 Investigating the measurement model variables

Before getting into the research hypotheses and conceptual model, it is necessary to ensure the measured variables' validity. Thus, we measure the exogenous and endogenous latent variables in the following research models discussed by first-order CFA (see Table 2). The CFA results suggest that our measurement model and variables provide an excellent fit to the data.

Table 2. CFA of the measurement model.

| Construct | Measure | Mean | Std dev | Loading factor |
|---|---|---|---|---|
| UCB awareness | 1. UCB banking is essential for the family. | 4.63 | 0.54 | 0.76 |
| | 2. I think others should be encouraged to store UCB. | 4.64 | 0.55 | 0.80 |
| | 3. I agree that UCB banking should be popularized. | 4.60 | 0.60 | 0.65 |
| Price | 4. The operation fee is reasonable. | 3.52 | 0.83 | 0.60 |
| | 5. Annual storage fee is reasonable. | 3.84 | 0.85 | 0.94 |
| Reference group | 6. Family's and friends' advice to store UCB has great importance. | 4.54 | 0.60 | 0.64 |
| | 7. Doctor's recommendation to store UCB has great importance. | 4.66 | 0.52 | 0.75 |
| | 8. "Media" suggestion to store UCB has great importance. | 4.79 | 0.42 | 0.70 |
| Usability | 9. UCB is viable biological insurance. | 4.49 | 0.59 | 0.64 |
| | 10. Own cord blood is thought to be safer than others. | 4.49 | 0.67 | 0.55 |
| | 11. The overall chance of using banked UCB for me is high compared with conventional treatments. | 3.80 | 0.79 | 0.55 |
| | 12. UCB stem cells can be used to treat many diseases and abnormalities my baby might encounter in his/her life. | 4.59 | 0.58 | 0.61 |
| | 13. UCB is useful for treatment. | 4.47 | 0.62 | 0.73 |
| Disease history | 14. Hematopoietic and nonhematopoietic diseases in the family have influenced my decision in UCB banking. | 4.12 | 0.94 | 0.73 |
| | 15. The anticipation of feeling guilty if the child needs cord blood, and it has not been stored. | 4.38 | 0.88 | 0.81 |
| | 16. I don't have a family history of the disease, but I am affluent enough for this service. | 4.23 | 0.85 | 0.77 |
| Perceived risk | 17. Cord blood banking may improve the health of my family. | 4.14 | 0.80 | 0.64 |
| | 18. I would participate in cord blood banking only if my relatives or friends do so. | 3.75 | 0.75 | 0.75 |
| | 19. I would participate in cord blood banking if cord blood banking technology were mature enough. | 4.21 | 0.82 | 0.77 |
| Behavioral intention | 20. I have a good image of cord blood banking. | 4.14 | 0.78 | 0.69 |
| | 21. I have specific plans to participate in cord blood banking. | 4.13 | 0.74 | 0.64 |
| | 22. I have the intention to use cord blood banking technology. | 4.26 | 0.76 | 0.80 |

### 4.3 Correlation analysis of the research variables

For testing the relationship between variables, we used Pearson correlation analysis, where the null hypothesis and hypothesis 1 are as follows:

$H_0$: The correlation coefficient is zero (no significant relationship between two variables).

$H_1$: The correlation coefficient is not zero (relationship between two variables).

As the results of Table 3 show, among the independent variables, usability has a significant relationship with perceived risk, and perceived risk has a significant relationship with behavioral intention.

9**Table 3.** Pearson correlation coefficient analysis

|  |  | UCB awareness | Price | Reference group | Usability | Disease history | Perceived risk | Behavioral intention |
|---|---|---|---|---|---|---|---|---|
| Awareness | Pearson correlation | 1 | .25** | .55** | .46** | .13* | -.09 | .02 |
|  | sig. (2-tailed) | --- | .00 | .00 | .00 | .04 | .19 | .80 |
| Price | Pearson correlation | .25** | 1 | .20** | .08 | -.01 | .01 | -.03 |
|  | sig. (2-tailed) | .00 | --- | .00 | .23 | .89 | .92 | .63 |
| Reference group | Pearson correlation | .55** | .20** | 1 | .44** | .19** | -.01 | .05 |
|  | sig. (2-tailed) | .00 | .00 | --- | .00 | .00 | .88 | .41 |
| Usability | Pearson correlation | .46** | .08 | .44** | 1 | .25** | -.24** | .02 |
|  | sig. (2-tailed) | .00 | .23 | .00 | --- | .00 | .00 | .76 |
| Disease history | Pearson correlation | .13* | -.01 | .19** | .25** | 1 | -.11 | .05 |
|  | sig. (2-tailed) | .04 | .89 | .00 | .00 | --- | .09 | .46 |
| Perceived risk | Pearson correlation | -.09 | .01 | -.01 | -.24** | -.11 | 1 | .23** |
|  | sig. (2-tailed) | .19 | .92 | .88 | .00 | .09 | --- | .00 |
| Behavioral intention | Pearson correlation | .02 | -.03 | .05 | .02 | .05 | .23** | 1 |
|  | sig. (2-tailed) | .79 | .63 | .41 | .76 | .46 | .00 | --- |

The sample size for all the analysis is $N=242$
* Correlation is significant at the 0.05 level (2-tailed).
** Correlation is significant at the 0.01 level (2-tailed).

### 4.4 Evaluation of structural research model (path analysis)

The null hypothesis and hypothesis 1 for verification or rejection are as follows:
$H_0$: There is no significant relationship between two variables.
$H_1$: There is a significant relationship between two variables.

If the test's significance value in the regression analysis is greater than 1.96, the null hypothesis is rejected, and hypothesis 1 is verified and vice versa. Table 4 shows the verification or rejection among research variables in summary.

**Table 4.** Structural equation modeling: Causal relationship analysis

| Research variables | Effect | Significance | Results |
|---|---|---|---|
| The effect of UCB awareness on perceived risk | 0.02 | 0.13 | Not supported |
| The effect of price on perceived risk | 0.02 | 0.22 | Not supported |
| The effect of the reference group on perceived risk | 0.23 | 1.54 | Not supported |
| The effect of usability on perceived risk | 0.41 | 2.76 | Supported |
| The effect of disease history on perceived risk | 0.04 | 0.48 | Not supported |
| The effect of perceived risk on behavioral intention | 0.30 | 3.34 | Supported |

In the test of the research hypotheses by structural equation model, the results demonstrate an acceptable model fit: $\chi^2 = 251.74$, DF = 193, RMSEA = 0.036, NFI = 0.90, NNFI = 0.96, PNFI = 0.75, CFI = 0.97, IFI = 0.97, RFI = 0.88, GFI = 0.91, and AGFI = 0.89. Fit statistics are all within the suggested threshold (Browne and Cudeck, 1992; Hu and Bentler, 1999). Based on the above statistical data, all SEM indicators were acceptable. These results reported in Table 5 confirmed that the conceptual model of perceived risk is an acceptable and appropriate theoretical model for examining the determinants of cord blood banking.

**Table 5.** Measurement item properties

| Research variables | Significance | T-value | Mean | Result |
|---|---|---|---|---|
| UCB awareness | .00 | 53.87 | 4.62 | Very high |
| Price | .00 | 14.17 | 3.68 | High |
| Reference group | .00 | 62.04 | 4.66 | Very high |
| Usability | .00 | 47.03 | 4.37 | Very high |
| Disease history | .00 | 25.58 | 4.24 | Very high |
| Perceived risk | .00 | 24.79 | 4.03 | Very high |
| Behavioral intention | .00 | 29.86 | 4.18 | Very high |

## 5 Discussion and conclusions

### 5.1 Summary of the results

The focus of this research is getting a clearer understanding of private UCB banking perceived risk determinants in Iran. We have examined the key factors that may impact consumers' perceived risk and behavioral intention in private cord blood banking.

The results show that 71% of the population has a university degree, and 26% have a high school or a diploma degree. These findings recommend that there is a relationship between educational level and UCB banking (Katz *et al.*, 2011; Yang *et al.*, 2003).

About 71% of the customers are from low-income-level families (less than $750), 22% of customers are from middle-income-level families (an income between $750 and $1500) and 7% have an income above $1500 per month. The results reveal that family income is not a determinant for private cord blood banking in Iran. In other words, the price of private cord blood banking was affordable so that even the families with lower income levels could take part. The finding of this research supports the study of Katz *et al.* (2011), which highlights that UCB banking choice is independent of income.

The empirical result of this research revealed the following significance. We classified them based on our hypotheses.

1. Hypothesis 1 of this research: UCB awareness did not reveal a negative influence on the perceived risk. Several studies in the past revolved around knowledge and attitude of pregnant women about UCB banking, such as in Canada and Turkey (Fernandez et al., 2003; Dinc and Sahin, 2009), and it has been found that women have a lack of knowledge about UCB banking (Huang, 2003). The sample population of the past studies was taken from pregnant women in hospitals; however, exceptionally, this research is focused on Royan cord blood bank customers who are already aware of UCB banking. The results show that people who have the intention to bank UCB in Royan are confidently aware of UCB applications. Another possible reason is that UCB banking awareness is not the primary source of perceived risk in Iran, and other factors such as usability, culture, and reputation are critical. Besides, this study's demographic results revealed that the majority of the respondents were educated, which could have been the reason for a high level of knowledge and UCB awareness.

2. Hypothesis 2 of this research: reference group did not reveal a negative influence on the perceived risk in UCB banking. This can be explained by mentioning the point that UCB banking is a socially accepted behavior in Iran, which has some of the most liberal laws on stem cell research in the world, and both the government and the society show a supportive attitude toward UCB banking for personal use. Although random sampling is used and past studies heavily focused on women's attitude (Fernandez *et al.*, 2003; Kharaboyan *et al.*, 2007; Dinc and Sahin, 2009), surprisingly, 85% of the respondents were men. A reason to explain this result is that in many developing countries such as Iran, men are the moneymakers, and they have the final word in decision making for the family. The role of fathers in the decision-making process in UCB banking has been studied by Katz *et al.* (2011), supporting our findings. The result is the key indicator that to commercialize cord blood banking, not only pregnant women but also fathers should be involved in getting access to appropriate information about UCB banking because they involve an important part of responsibility in deciding whether to bank their child's cord blood.

3. Hypothesis 3 of this research: the empirical findings of this research revealed a significant and negative influence of usability on UCB banking perceived risk. Besides, in the past, the studies reported by several researchers exhibit noticeable uncertainty about UCB banking usability for personal use (Fisk *et al.*, 2005). Private cord blood banking has been criticized by giving unreal hopes to future parents to save UCB stem cells with very little chance of getting used in the future (Brown, 2007; Fox *et al.*, 2007). In line with the literature, the negative influence of usability on perceived risk is verified in this research.

4. Hypothesis 4 of this research: price did not reveal a positive influence on UCB banking perceived risk. According to literature, the price was a barrier to entry of people to private cord blood banks (Katz *et al.*, 2011), but in Iran, price is not a determinant for UCB banking. The main reason was the low cost of private UCB banking in Iran, which is roughly $500 for the operation fee and $40 for the annual fee; however, in developed countries such as the United States, private cord blood banking cost ranges from $1000 to $2000 for operation fee and approximately $100 to $200 for an annual fee. Accordingly, we can find out that in a low-cost private cord blood bank, price is not a determining factor of the respondent's choice.

5. Hypothesis 5 of this research: disease history did not reveal a negative influence on UCB banking perceived risk, which responded to the fact that having a disease history in a family would not promote the feasibility of UCB banking for personal use in Iran. This can be explained by pointing out that the price people pay to bank their child's UCB stem cells in Iran compared with that of developed countries is almost half. For that reason, for a family to decide on banking their child's UCB stem cells, having a disease history will not play a role. People in Iran are affluent to privately save their child's UCB stem cells, which is highly commercialized as a

12"once-in-a-lifetime opportunity" and "biological insurance" that may save their child's life in the future.

6. Hypothesis 6 of this research: perceived risk revealed significant and negative influence on intention to bank UCB. This is because private UCB banking is a high-risk service that is not a necessity for families. This result is in line with several studies, such as the theory of perceived risk suggested by Taylor (1974) and Lin (2008), which indicates the significant effect of perceived risk on the intention to purchase a product or use a service.

### 5.2   Implications for managers

Whether and why consumers will adopt a new medical service should be well understood as a vital insight for managers concerned with marketing innovations. To alleviate perceived risk, in particular, cord blood banks must be directly addressed. The result indicates new paths for future explorations and provides insights to UCB banking administrators, and can be incorporated into government regulatory policy. There is increasing literature focusing on the economization of human biological life, and this case study will make a significant contribution to this. The usability of UCB stem cells is the leading cause of consumers' perceived risk. Therefore, to promote UCB banking adoption, service must start from improving the usability and reducing the perceived risk.

### 5.3   Limitations and future research

This research inevitably has limitations. The sample population of this research was taken from Royan cord blood bank customers, and the respondents already had knowledge and awareness about UCB banking for personal use. This paper creates an avenue for future studies, especially in the gray areas identified. The cross-regional samples in future studies may move on to examine whether the customers in developed countries perceive UCB banking differently from those in Iran and other developing countries. The storage of UCB should not be treated as a commercial activity and instead as an intention to advance the development of stem cell research for a safe and healthy life. As we tried to keep the analysis as simple as possible, more advanced statistical techniques such as robust estimation of multivariate processes, which is not sensitive to outlier data, as applied in Ebadi et al. (2011), can also be implemented.

## References

trueAaker, D. (1991), *Managing Brand Equity*, Free Press, New York, NY, pp. 19–32.

Armson, B. A., Allan, D. S., & Casper, R. F. (2015). Umbilical cord blood: counseling, collection, and banking. *Journal of Obstetrics and Gynaecology Canada*, *37*(9), 832-844.

Barkhordari, A., Malmir, B., & Malakoutikhah, M. (2019). An analysis of individual and social factors affecting occupational accidents. *Safety and health at work*, *10*(2), 205-212.